\documentstyle[preprint,aps,eclepsf]{revtex}
\newcommand{\postscript}[1]{\hbox{\epsfile{file=#1}}} 
\begin{document}
\begin{center}
{\Large \bf Method of constructing  
   exactly solvable chaos}\footnote{To appear in {\it Phys. Rev.}{\bf E}(1997)}\\
\vskip.25in
{ Ken Umeno}
  
{\it Frontier Research Program,
 The Institute of Physical 
  and Chemical Research (RIKEN)\\ 
2-1 Hirosawa, Wako, Saitama 351-01, Japan} 
\vskip.25in
\end{center}

\begin{abstract}
We present a new systematic method of  constructing  rational mappings
as ergordic transformations with nonuniform 
invariant
measures on the unit interval \(\mbox{\boldmath$I$}=\left[0,1\right]\).
As a result, we obtain a two-parameter family of rational mappings that 
have a special property in that their invariant measures can be 
explicitly written in terms of algebraic functions of parameters and 
a dynamical variable.  
Furthermore, it is  shown  here 
that this  
family is the most generalized class of rational mappings possessing 
the property of 
exactly solvable chaos 
on \(\mbox{\boldmath$I$}\), 
including the Ulam=Neumann map \(y=4x(1-x)\). 
Based on the present method, we can produce a series of   
rational mappings resembling    
 the asymmetric shape of   
the experimentally obtained first return maps 
of the Beloussof-Zhabotinski 
chemical reaction, and  we can match some rational functions with 
 other experimentally  obtained 
first return maps in a systematic manner. 
\\05.45+b, 04.20.Jb
\end{abstract}
\vskip.25in
\clearpage 
\setcounter{equation}{0} 
Characterizing invariant measures for explicit nonlinear dynamical systems 
 is a fundamental problem which 
connects  dynamical theory with statistics and  
statistical mechanics. In some cases, it would be desirable to 
to characterize ergordic invariant measures for simple chaotic 
dynamical systems. 
However, in the cases of chaotic dynamical systems, such  
attempts to obtain explicit invariant measures have rarely been  made.  
One well-known exception is the logistic map 
\(Y=4X(1-X)\equiv f_{0}(X)\) on 
\(\mbox{\boldmath$I$}=\left[0,1\right]\)  given 
by Ulam and von Neumann in the late 1940's\cite{uf}. 

The Ulam=Neumann dynamical system 
\(x_{i}=f_{0}(x_{i-1})\)
has an ergordic measure \(\mu(dx)=\frac{dx}{\pi\sqrt{x(1-x)}}\) 
such that the time 
averages of a  function \(Q(x)\) 
can be explicitly computed by the formula  
\(
  \lim_{N\rightarrow \infty}
\sum_{i=0}^{N-1}\frac{1}{N}Q(x_{i}) = 
\int_{0}^{1}\frac{Q(x)dx}{\pi\sqrt{x(1-x)}} 
\)
for almost all initial conditions \(x_{0}\in \mbox{\boldmath$I$}\). 
The first attempt to generalize the 
Ulam=Neumann map within a set of rational functions was made by 
Katsura and Fukuda in 1985\cite{kf}.  The Katsura and Fukuda model is 
written as  
\begin{equation}
\label{eq:katsura}
Y=\frac{4X(1-X)(1-lX)}{(1-lX^{2})^{2}}\equiv f_{l}(X)  
\end{equation} 
 for \(0\leq l < 1\). Clearly, 
the Ulam=Neumann map can be regarded as  a special 
case of  
Katsura=Fukuda systems where  the parameter \(l\) is set to \(0\). 
The  author  showed  \cite{ku2} 
that the Katsura=Fukuda mappings (\ref{eq:katsura}) also have ergordic 
measures which can be written explicitly as 
\begin{equation}
\label{eq:kfi}
\mu(dx)=\rho(x)dx= \frac{dx}{2K(l)\sqrt{x(1-x)(1-lx)}},
\end{equation} 
where \(K(l)\) is the elliptic integral of the first kind \((g=1)\) given by 
\(
  K(l) = \int_{0}^{1} \frac{du}{\sqrt{(1-u^{2})(1-lu^{2})}}
\).
It is known \cite{kf} that  
the Katsura=Fukuda  systems and  the Ulam=Neumann system 
also have explicit solutions in terms of the Jacobi \(sn\) elliptic function, 
as 
\begin{equation}
\label{eq:expression1}
  x_{n}= sn^{2}(K(l)2^{n}\theta_{0}), \quad \theta_{0}\in \mbox{\boldmath$I$},
\end{equation} 
 where \(\sqrt{l}\) corresponds to the modulus of Jacobi elliptic functions.
The validity of the formulae of the 
 general solutions (\ref{eq:expression1}) is easily checked  
 using the duplication 
formula\cite{ww} of the Jacobi \(sn\) elliptic function  
\begin{equation}
sn(2u)=\frac{2sn(u)\sqrt{(1-sn^{2}(u))(1-lsn^{2}(u))}}{(1-lsn^{4}(u))}.
\end{equation} 
Because  the Ulam=Neumann system and Katsura=Fukuda systems have 
{\it not only}  exact  
solutions (\ref{eq:expression1}) {\it but also}  
 explicitly written ergordic invariant measures (\ref{eq:kfi}), 
 we call a dynamical system 
{\it exactly solvable chaos} if it has 
both of these properties.
 
Thus, it is of great interest to investigate whether  
we can  generalize the Ulam=Neumann  system and the 
Katsura=Fukuda systems further within a set of rational functions,  
maintaining the property of exactly solvable chaos. 
The main purpose of the present letter is to show that by using 
 the   addition formulas of elliptic functions we can construct  
  a two-parameter family of rational mappings of exactly solvable chaos, 
and at the same time, that  there is 
a certain limitation to  
generalizing this family within a set of rational functions.

Our results reported here \cite{ku0} 
concern the following rational  transformations  
\begin{equation}
\label{eq:formula1}
Y=f_{l,m}(X)=\frac{4X(1-X)(1-lX)(1-mX)}{1+AX^{2}+BX^{3}+CX^{4}}\in I,
\end{equation}
  where 
\(A=-2(l+m+lm), 
 B=8 lm,
C= l^{2}+m^{2}-2lm-2l^{2}m-2lm^{2}+l^{2}m^{2}\), and 
\(X\in I\). 
The parameters \(l\) and \(m\) are arbitrary real numbers 
satisfying the condition 
 \(-\infty <m\leq l<1\). 
Figure 1(a) shows various shapes of the proposed  mappings (\ref{eq:formula1}).
Surprisingly,  
some rational maps in (\ref{eq:formula1}) strongly  
resemble the  asymmetric shape 
of the experimentally obtained first return maps  
of the Beloussof-Zhabotinski chemical reaction. 
Here, we will prove the following statement. The two-parameter family 
of rational  mappings (\ref{eq:formula1}) 
is also exactly solvable chaos such that the dynamical systems 
\(x_{i+1}=f_{l,m}(x_{i})\)  (\ref{eq:formula1}) have 
ergordic invariant measures
explicitly given by 
\begin{equation}
  \mu(dx)=\rho(x)dx=
  \frac{dx}{2K(l,m)\sqrt{x(1-x)(1-lx)(1-mx)}},
\end{equation}
 where \(K\) is given by the   
 integrals  
\begin{equation}
  K(l,m)=\int_{0}^{1}\frac{du}{\sqrt{(1-u^{2})(1-lu^{2})
   (1-mu^{2})}},
\end{equation}
and it has  general solutions explicitly given by  
\begin{equation}
  x_{n}=s^{2}(K(l,m)2^{n}\theta_{0})\quad \theta_{0}\in \mbox{\boldmath$I$}.
\end{equation}
We prove this  by explicitly computing  the duplication formula
of the following {\it degenerated} hyperelliptic function \(s(x)\) defined 
by  
\begin{equation}
\label{eq:degenerate}
  s^{-1}(x)=\int_{0}^{x}\frac{du}{\sqrt{(1-u^{2})(1-lu^{2})
   (1-mu^{2})}}.
\end{equation} 
Here, {\it degenerated}  means that 
this hyperelliptic integral of the R.H.S. of Eq. (\ref{eq:degenerate}) can be
reduced to a certain elliptic integral by a rational change of variables. 
Although the  reduction of Abelian integrals of genus \(g\geq 2\) to 
elliptic functions
was intensively studied in the 19th century by Jacobi, 
Weierstrass, K\"onigsberger, 
 Kovalevskaya and others, it was only in the 1980's
 that  
 the theory of the reduction was successfully applied to physics for   
obtaining, for example,  explicit periodic solutions in terms of elliptic 
functions for
 the  Korteweg-de Vries (KdV) equation and the sin-Gordon equation\cite{bbme}.
Let us consider  reduction of the hyperelliptic integrals 
(\ref{eq:degenerate}) as
\begin{equation}
\label{eq:reduction}
 s^{-1}(x)=\int_{0}^{x}\frac{du}{\sqrt{(1-u^{2})(1-lu^{2})
    (1-mu^{2})}}=\int_{0}^{x^{2}}\frac{dv}{2\sqrt{v(1-v)(1-lv)(1-mv)}},
\end{equation}
 where \(u^{2}=v\). 
Thus, we can write 
the degenerated hyperelliptic function \(s(x)\) 
in terms of the Weierstrass elliptic 
functions. 
The Weierstrass elliptic function \(\wp(u)\) of 
\(u\in \mbox{\boldmath$C$}\) is defined by 
\begin{equation}
  \wp(u)=\frac{1}{u^{2}}+{\sum_{j,k}}^{'}\{
\frac{1}{(u-2j\omega_{1}-2k\omega_{2})^{2}}-
\frac{1}{(2j\omega_{1}+2k\omega_{2})^{2}}\},
\end{equation}
  where the symbol \(\sum'\) means that the summation is made over 
all combinations of integers \(j\) and \(k\), except for the combination 
\(j=k=0\), and  \(2\omega_{1}\) and \(2\omega_{2}\) are  periods of 
the function \(\wp(u)\)\cite{ww}. The Weierstrass elliptic function 
\(\wp(u)\) 
satisfies the differential equation
\begin{equation}
\label{eq:we}
  (\frac{d\wp (x)}{dx})^{2}=4\wp^{3}(x)-g_{2}\wp(x)-g_{3},
\end{equation}
 with the invariants   
  \(g_{2}(\omega_{1},\omega_{2}) =
 60{\sum_{j,k}^{'}}\frac{1}{(j\omega_{1}+k\omega_{2})^{4}}\) and  
   \(g_{3}(\omega_{1},\omega_{2})=
140{\sum_{j,k}^{'}}\frac{1}{(j\omega_{1}+k\omega_{2})^{6}}\)\cite{ww}. 
Let \(e_{1},e_{2}\) and \(e_{3}\) be the roots of the equation 
\(4z^{3}-g_{2}z-g_{3}=0\); that is,
\begin{equation}
  e_{1}+e_{2}+e_{3}=0,\quad 
e_{1}e_{2}+e_{2}e_{3}+e_{3}e_{1}=-\frac{g_{2}}{4},\quad 
e_{1}e_{2}e_{3}=\frac{g_{3}}{4}. 
\end{equation}
The {\it discriminant} \(\Delta\) of the function \(\wp(u)\) is given 
by \(\Delta=g_{2}^{3}-27g_{3}^{2}\). If \(\Delta>0\), all roots 
\(e_{1},e_{2}\) and \(e_{3}\) of the equation \(4z^{3}-g_{2}z-g_{3}=0\) 
are {\it real}. Thus, we can assume that \(e_{1}>e_{2}>e_{3}\). 
In the case that \(\Delta>0\), it is known
that the periods \(\omega_{1}\) and \(\omega_{2}\) are  written simply as 
\begin{equation}
\label{eq:omegaint}
\omega_{1}=\int_{e_{1}}^{\infty}\frac{dz}{\sqrt{4z^{3}-g_{2}z-g_{3}}},
\quad 
\omega_{2}=i\int_{-\infty}^{e_{3}}\frac{dz}{\sqrt{g_{3}+g_{2}z-4z^{3}}}.
\end{equation} 
Using  
the transformation of the variable as 
   \(v=-\frac{(1-l)(1-m)}{y-\frac{2l+2m-3lm-1}{3}}+1\),
we can rewrite \(s^{-1}(x)\)  in Eq.(\ref{eq:reduction}) 
as 
\begin{equation}
\int_{0}^{x^{2}}\frac{dv}{2\sqrt{v(1-v)(1-lv)(1-mv)}}=
\int_{\frac{2-l-m}{3}}^{\frac{2l+2m-3lm-1}{3}+\frac{(1-l)(1-m)}{1-x^{2}}}
     \frac{dy}{\sqrt{4y^{3}-g_{2}y-g_{3}}},
\end{equation}
where
 \(g_{2}=\frac{4(1-l+l^{2}-m+m^{2}-lm)}{3}\) and 
\(g_{3}=\frac{4(2-l-m)(2l-m-1)(2m-l-1)}{27}\). 
We note here that \(4y^{3}-g_{2}y-g_{3}\) can be factored as 
\begin{equation}
  4y^{3}-g_{2}y-g_{3}=4(y-\frac{2-l-m}{3})(y-\frac{2l-m-1}{3})
(y-\frac{2m-l-1}{3}).
\end{equation} 
We  set
\( e_{1}=\frac{2-l-m}{3}>e_{2}=\frac{2l-m-1}{3}>e_{3}=\frac{2m-l-1}{3}\).
 Thus, 
using the integral representation of the period
\(\omega_{1}\) (\ref{eq:omegaint})
 and the differential equation (\ref{eq:we}) for 
the Weierstrass elliptic function,
\(s(x)\) can be written explicitly 
in terms of the Weierstrass elliptic function as
\begin{equation}
\label{eq:rational}
 s^{2}(x)=1-\frac{(1-l)(1-m)}{\wp(\omega_{1}-x)-\frac{2l+2m-3lm-1}{3}}. 
\end{equation}
The function  \(s^{2}(x)\) also has  the same periods  
\(\omega_{1}\) and 
\(\omega_{2}\) computed using  formula (\ref{eq:omegaint}). 
It is noted here that because 
\begin{equation}
  \Delta\equiv g^{3}_{2}-27g^{2}_{2}=16(1-l)^{2}(1-m)^{2}
  (l-m)^{2}>0  
\end{equation} 
for \(-\infty <m<l<1\), the period \(2\omega_{1}\) is always {\it real}, 
while 
the period \(2\omega_{2}\) is always {\it pure imaginary}.
Using
the addition formula,
\begin{equation}
  \wp(z+y)=\frac{1}{4}\{\frac{\wp'(z)-\wp'(y)}{\wp(z)-\wp(y)}\}^{2}
  -\wp(z)-\wp(y),
\end{equation}
and 
the duplication formula,
\begin{equation}
  \wp(2z)=\frac{1}{4}\{\frac{\wp''(z)}{\wp'(z)}\}^{2}
   -2\wp(z),
\end{equation}
 for the Weierstrass elliptic function \(\wp(u)\) \cite{ww},
 we finally  obtain the {\it explicit} duplication formula of \(s(x)\) as 
\begin{equation}
\label{eq:duplication}
s^{2}(2x)=\frac{4s^{2}(x)(1-s^{2}(x))(1-ls^{2}(x))(1-ms^{2}(x))}
{1+As^{4}(x)+Bs^{6}(x)+Cs^{8}(x)},
\end{equation}
 where 
\(A=-2(l+m+lm),
 B=8 lm\) and 
\(C= l^{2}+m^{2}-2lm-2l^{2}m-2lm^{2}+l^{2}m^{2}\).
If we set \(X=s^{2}(x)\) and \(Y=s^{2}(2x)\), we obtain system 
(\ref{eq:formula1}) as  
\(Y=f_{l,m}(X)\). 
 
Using the relations  
\begin{equation}
  s^{2}(\omega_{1}\cdot 2\theta)=f_{l,m}(s^{2}(\omega_{1}\theta)),\quad
  s^{2}(\omega_{1}\cdot (2-2\theta))=f_{l,m}(s^{2}(\omega_{1}\theta)),
\end{equation} 
 for \(\theta \in \left[0,1\right]\)
and by defining the homeomorphism of \(\left[0,1\right]\) into itself 
given by 
  \(\phi_{l,m}(x)=\frac{1}{\omega_{1}}s^{-1}(\sqrt{x})\), 
 we derive the 
 tent map \(\tilde f(x)=\phi_{l,m}\circ f_{l,m} \circ\phi_{l,m}^{-1}\) as  
\begin{equation}
\label{eq:tent}
   \tilde f(x) = 2x\quad \mbox{for}\quad  
x\in\left[0,\frac{1}{2}\right], \quad 
    \tilde f(x)=2-2x\quad \mbox{for}\quad 
    x\in\left[\frac{1}{2},1\right]. 
\end{equation}
Because this tent map (\ref{eq:tent}) is clearly ergordic and  
preserves the Lebesgue measure, 
 the map \(f_{l,m}\) preserves the  
measures
\begin{equation}
\label{eq:invariantm}
  \mu(dx)=\frac{d\phi_{l,m}}{dx}dx=
\frac{dx}{2K(l,m)\sqrt{x(1-x)(1-lx)(1-mx)}}. 
\end{equation} 
 This measure (\ref{eq:invariantm}) is  
  absolutely continuous 
   with respect to the Lebsegue measure, which implies that 
  the Kolmogorov-Sinai entropy \(h(\mu)\) is equivalent 
 to the Lyapunov exponent 
  of \(\log 2\) from the Pesin identity \cite{eckmann}, and that the measure 
  (\ref{eq:invariantm}) is a physical one in the sense that it is 
  the Sinai-Ruelle-Bowen (SRB) measure such that  
 for almost all initial 
conditions \(x_{0}\), the time averages
    \(\lim_{N\rightarrow \infty} 
\frac{1}{N}\sum_{i=0}^{N-1}\delta (x-x_{i})\)
reproduce the invariant measure \(\mu(dx)\)\cite{eckmann}.
 
In the same way, we can construct 
{\it generalized cubic maps} \(f^{(3)}_{l,m}\) 
 from the triplication formula 
\(s^{2}(3x)=f^{(3)}_{l,m}(s^{2}(x))\) as
\begin{equation}
\label{eq:gcubic}
Y=f^{(3)}_{l,m}(X)=
\frac{X(-3+4X+\sum_{i=1}^{4}A_{i}X^{i})^{2}}
{1+\sum_{i=2}^{9}B_{i}X^{i}},
\end{equation}
 where \(A_{1},\cdots,A_{4}\) and \(B_{2},\cdots,B_{9}\) 
are polynomial functions in the parameters \(l\) and \(m\) 
which vanish for \(l=m=0\)\cite{precise}. 
The generalized cubic map \(f^{(3)}_{l,m}\) has 
the same invariant measures (\ref{eq:invariantm}) because  the relation 
\(\tilde f^{(3)}(x)=\phi_{l,m}\circ f^{(3)}_{l,m}\circ \phi^{-1}_{l,m}\)   
holds for the piecewise-linear map 
\(\tilde f^{(3)}(x)=3x\) for \(0\leq x \leq \frac{1}{3}\), \(-3x+2\) for 
    \(\frac{1}{3} \leq x \leq \frac{2}{3}\) and 
  \(3x-2\) for \(\frac{2}{3} \leq x \leq 1\).
If we set \(l=m=0\), this rational mapping is reduced to the cubic map 
\(Y=X(3-4X)^{2}\) 
as a special case of 
Chebyshev maps obtained by Adler and Rivlin\cite{adler}.  
Thus, we can obtain {\it generalized Chebyshev maps} as rational functions 
\(f^{(p)}_{l,m}\) 
from the addition formulas \(s^{2}(px)=f^{(p)}_{l,m}(s^{2}(x))\), 
which have the same invariant
 measures (\ref{eq:invariantm}).       
The shapes of generalized cubic maps are depicted in Fig. 1(b). 
 Based on the case of Beloussof-Zhabotinski map,  we  predict here that 
 some rational mappings (\ref{eq:gcubic}), such as that shown in  Fig. 1(b), 
 can resemble the  first return maps experimentally 
constructed from some 
 unknown  chaotic phenomena. 

Are there more generalized rational mappings that possess
the   properties of exactly solvable chaos such as  
(\ref{eq:formula1})? 
The exact solvability of the present rational mappings (\ref{eq:formula1})
and (\ref{eq:gcubic}) is 
due to the fact that \(s^{2}(x)\) in Eq. (\ref{eq:rational})
is a rational function of the Weierstrass elliptic function \(\wp(u)\)
having  the {\it addition  formula}  and  
 the real period 
\(\omega_{1}\).  
For an arbitrary 
 set of parameters \(e_{1},e_{2}(<e_{1})\), and \( e_{3}(=-e_{1}-e_{2})\) 
which determine   
the Weierstrass elliptic function \(\wp(u)\) with the real period, 
there exists a set of parameters 
\(l\) and \(m\) of (\ref{eq:formula1}) given by 
\(l=1-(e_{1}-e_{2})\) and \(m=1-(2e_{1}-2e_{2})<l\); i.e., the mapping 
 \(h: (e_{1},e_{2})\mapsto 
  (l,m)\)  
 is a {\it bijection}. In other words, the family \(\{f_{l,m}\}\) 
has one-to-one correspondence  
 to the set of  
Weierstrass elliptic functions \(\{ \wp (u) \}\) with real periods.  
Furthermore, any elliptic function \(w(u)\) can be expressed  
in terms of Weierstrassian elliptic 
functions \(\wp(u)\) and \(\wp(u)'\) with the same periods, the expression 
being rational in \(\wp(u)\) and linear in \(\wp'(u)\)\cite{ww,ww1}. 
Since it is known \cite{hc} from the {\it Weierstrassian theorem} that 
the  class of functions \(h(u)\) having the algebraic addition  formulae
\cite{aaddition} including 
the duplication formulae is exactly 
 the class of algebraic functions of elliptic functions \(\wp(u)\), which 
of course includes the class of algebraic functions of  sin functions,
 we can say  that 
 this two-parameter family of dynamical systems  
 essentially forms a maximal class in representing 
exactly solvable chaos 
 induced by one-dimensional   
rational mappings.  

In conclusion, 
we present a new method of constructing 
 ergordic transformations related to rational functions with 
explicit nonuniform 
invariant measures using  the addition formulas of elliptic functions. 
As a result, we 
systematically generalize the Ulam=Neumann (logistic) map and 
Chebyshev maps into 
two-parameter families of rational mappings.
 We also showed that 
constructing the more generalized  family of 
rational mappings possessing explicit ergordic invariant 
measures  
 has a certain limitation due to  the  Weierstrassian 
argument concerning the addition  
formulas for general algebraic functions. 
As for the applications, 
all of the  constructive  results given here 
for ergordic invariant measures by rational mappings 
can be directly used as nonlinear 
random number generators for the Monte Carlo methods.  

The author  would like to acknowledge Dr. K. Iguchi and  Dr. A. Bobenko  for 
useful discussions. 
Also, Professor T. Kohda and Professor S. Amari provided helpful suggestions.
This work was supported by the RIKEN Special Researcher's Program to 
promote basic sciences.  
\begin{figure}[htb]
\begin{minipage}[t]{8.0cm}
\postscript{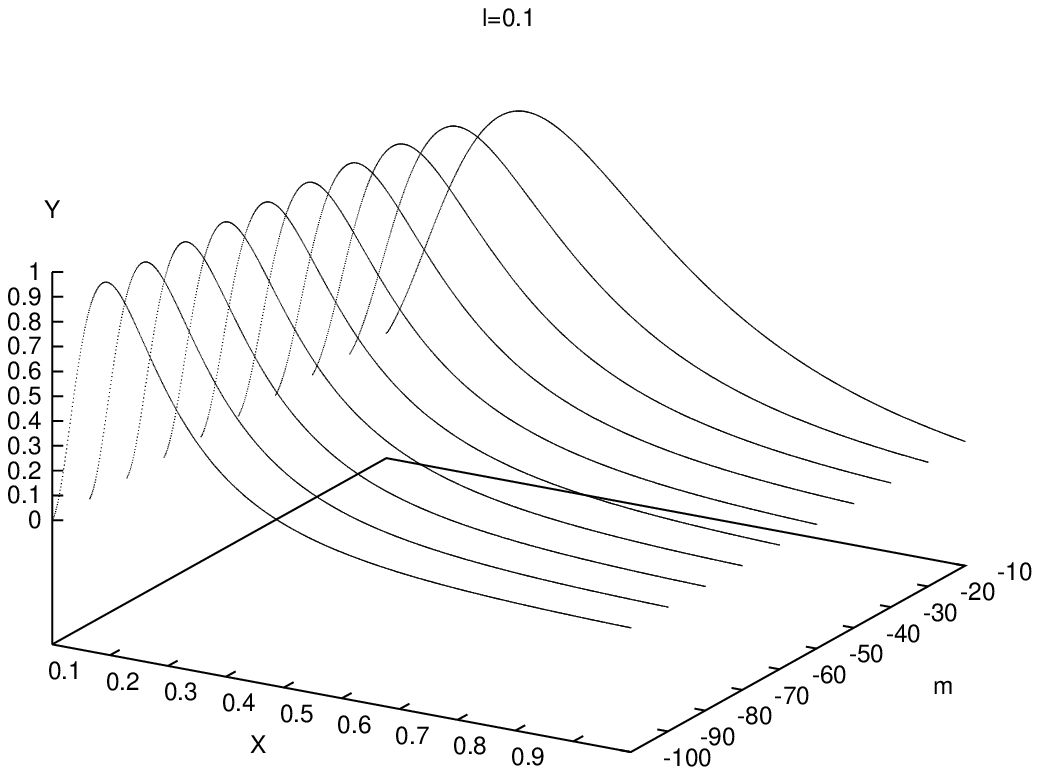,height=10cm}
\end{minipage} \hfill \\
{{\bf FIG. 1(a):}\ (mono).\quad
  The two-parameter family of exactly solvable chaos mappings
(\ref{eq:formula1})
for \((l,m)=(0.1,-10),(0.1,-20),\cdots(0.1,-100)\). The asymmetric shapes of
this class of mappings are very similar to the first return maps of the
Beloussof-Zhabotinski chemical reaction.}
\end{figure}
\begin{figure}[htb]
\begin{minipage}[t]{8.0cm}
\postscript{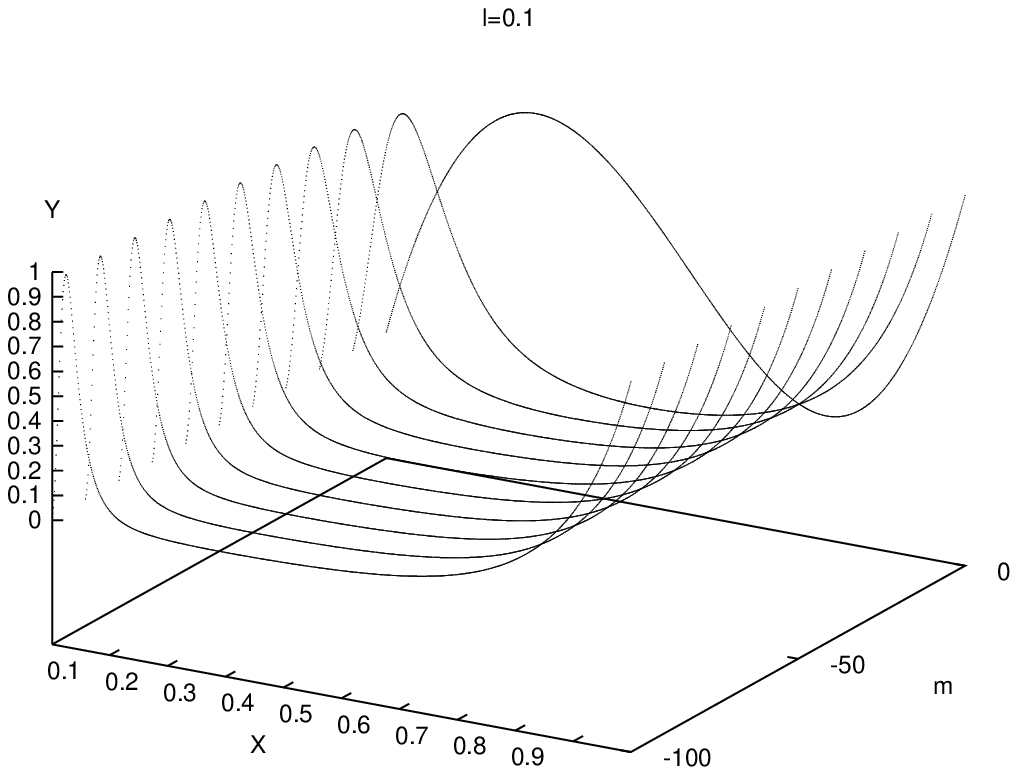,height=10cm}
\end{minipage} \hfill \\
{{\bf FIG. 1(b):}\ (mono).\quad
The two-parameter family of exactly solvable chaos mappings
(\ref{eq:gcubic})
for \((l,m)=(0.1,-10),(0.1,-20),\cdots(0.1,-100)\).
}
\end{figure}



\begin{thebibliography}{99}
\bibitem{precise}
 Precise forms of \(A_{1},\cdots,A_{4}\) and 
  \(B_{1},\cdots,B_{9}\) are given as 
\begin{eqnarray} 
\begin{array}{l}
A_{1}=4(l+m),A_{2}=-6(l+m+lm),
A_{3}=12lm,\\
A_{4}=l^{2}+m^{2}-2lm-2l^{2}m-2lm^{2}+l^{2}m^{2},\\
B_{2}=-12(l+m+lm),\\
B_{3}=8(l+m+l^{2}+m^{2}+l^{2}m+lm^{2}+15lm),\\
B_{4}=6(5l^{2}+5m^{2}-26lm-26l^{2}m-26lm^{2}+5l^{2}m^{2}),\\
B_{5}=24(-2l^{2}-2m^{2}-2l^{3}-2m^{3}+4lm+7l^{2}m+7lm^{2})\\
   +24(4l^{3}m+4lm^{3}+7l^{2}m^{2}-2l^{3}m^{2}-2l^{2}m^{3}),\\
B_{6}=4(4l^{2}+4m^{2}+4l^{4}+4m^{4}+17l^{3}+17m^{3}-8lm)\\
     +4(-17l^{2}m-17lm^{2}-17l^{3}m-17lm^{3}-8l^{4}m-8lm^{4})\\
     +4(4l^{2}m^{4}+4l^{4}m^{2}-17l^{3}m^{2}-17l^{2}m^{3}+17l^{3}m^{3}-
54l^{2}m^{2}),\\
B_{7}=24(-l^{3}-m^{3}-l^{4}-m^{4}+l^{2}m+lm^{2}-l^{3}m-lm^{3})\\
     +24(l^{4}m+lm^{4}+4l^{2}m^{2}+4l^{3}m^{2}+4l^{2}m^{3})\\
     +24(l^{4}m^{2}+l^{2}m^{4}-l^{3}m^{3}-l^{4}m^{3}-l^{3}m^{4}),\\
B_{8}=3(3l^{4}+3m^{4}+4l^{3}m+4lm^{3}+4l^{4}m+4lm^{4}-14l^{2}m^{2})\\
  +3(-4l^{3}m^{2}-4l^{2}m^{3}-4l^{3}m^{3}-14l^{4}m^{2}-14l^{2}m^{4}+
4l^{4}m^{3}+4l^{3}m^{4}+3l^{4}m^{4}
),\\
B_{9}=8(-l^{4}m-lm^{4}+l^{3}m^{2}+l^{2}m^{3}+l^{4}m^{2}+l^{2}m^{4}
      -2l^{3}m^{3}+l^{4}m^{3}+l^{3}m^{4}-l^{4}m^{4}).
\end{array}\nonumber
\end{eqnarray}
\bibitem{ww1} Thus, any elliptic function \(w(u)\)  has an algebraic relation 
 with the Weierstrass elliptic function \(\wp (u)\) with the same 
periods as
  \(G(w,\wp)=0\),   
\(G\) being  a rational function of two variables.
\bibitem{aaddition} We say that a function \(w(u)\) has an algebraic addition
formula if there exists a polynomial function \(G\) of three variables
such that the relation
 \(G(w(u_{1}+u_{2}),w(u_{1}),w(u_{2}))=0\) holds for
an arbitrary set of the variables \(u_{1}\) and \(u_{2}\).
\end{thebibliography}
\end{document}